\documentclass[11pt]{article}
\usepackage{amssymb,latexsym,amsmath, amsthm}
\usepackage{amsfonts, graphics,color}
\usepackage{mathtools}
\usepackage{hyperref} 
\usepackage{breakurl} 
\usepackage[american]{babel}
\usepackage{csquotes}

\usepackage[top=1in, bottom=1.25in, left=1.25in, right=1.25in]{geometry}

\usepackage{subcaption} 
\usepackage{float}

\def\R{\mathbb R}

\let\originalleft\left
\let\originalright\right
\renewcommand{\left}{\mathopen{}\mathclose\bgroup\originalleft}
\renewcommand{\right}{\aftergroup\egroup\originalright}
\newcommand{\ft}[0]{\footnotesize}
\newcommand{\sm}[0]{\normalsize}

\usepackage{color}
\usepackage[normalem]{ulem} 
\def\bcr{\begin{color}{red}}
\def\bcb{\begin{color}{blue}}
\definecolor{darkgreen}{RGB}{0,150,0}
\def\bcg{\begin{color}{darkgreen}}
\def\ec{\end{color}}
\def\be{\begin{equation}}
\def\ee{\end{equation}}

\hypersetup{pdftitle={GSR-EVstab}}

\theoremstyle{remark}

\numberwithin{equation}{section}

\title{Collisionless equilibria in general relativity: stable configurations beyond the first binding energy maximum}

\author{Sebastian G\"unther, Christopher Straub, Gerhard Rein \vspace{0.4cm}   \\ 
  Department of Mathematics, University of Bayreuth, Germany}

\begin{document}

\maketitle

\vspace{-.5cm}

\begin{abstract}
We numerically study the stability of collisionless equilibria in the context of general relativity. More precisely, we consider the spherically symmetric, asymptotically flat Einstein-Vlasov system in Schwarzschild and in maximal areal coordinates. Our results provide strong evidence against the well-known binding energy hypothesis which states that the first local maximum of the binding energy along a sequence of isotropic steady states signals the onset of instability. 
We do however confirm the conjecture that steady states are stable at least up to the first local maximum of the binding energy. For the first time, we observe multiple stability changes  for certain models. The equations of state used are piecewise linear functions of the particle energy and provide a rich variety of different equilibria. 
\end{abstract}

\tableofcontents

\section{Introduction}\label{sc:intro}
\subsection{The Einstein-Vlasov system}\label{ssc:evsystem}

In mathematical physics, the Einstein-Vlasov system is used to model galaxies or globular clusters in a relativistic setting. It describes how a large number of mass points behave which interact only through the Einstein equations
\begin{align} \label{eq:feqgen}
	G_{\alpha \beta} = 8 \pi T_{\alpha \beta}.
\end{align}
Here $G_{\alpha \beta}$ is the Einstein tensor induced by a Lorentzian metric $g_{\alpha \beta}$  with signature \mbox{$(-+++)$}, and $ T_{\alpha \beta}$ is the energy-momentum tensor. Greek indices always run from $0$ to $3$. Throughout this work, we set the speed of light and the gravitational constant to unity. 

By assuming that all particles have the same rest mass equal to $1$ and move forward in time, their number density $f$ can be written as a function $f=f(t,x^i,p^j)$ where Latin indices range from $1$ to $3$; we assume that $p^0$ can be expressed in terms of the other variables. The local coordinates in spacetime are denoted as \mbox{$(t,x^i)$}, and $p^\alpha$ are the corresponding canonical momentum variables. The evolution of the matter is then determined by the collisionless Boltzmann or Vlasov equation
\begin{equation}\label{eq:vlasovgeneral}
	\partial_t f + \frac{p^i}{p^0} \partial_{x^i} f - \frac{1}{p^0} \Gamma^i_{\beta \gamma} p^\beta p^\gamma \partial_{p^i} f = 0,
\end{equation}
where $\Gamma^\alpha_{\beta \gamma}$ are the Christoffel symbols induced by the metric. Note that the Einstein summation convention is applied. Einstein's equations are coupled to the Vlasov equation via the energy momentum tensor 
\begin{equation}\label{eq:energymomentum}
	T_{\alpha \beta} = \int_{\R^3} p_\alpha p_\beta f |g|^\frac{1}{2} \frac{dp^1dp^2dp^3}{-p_0},
\end{equation}
where $|g|$ is the modulus of the determinant of the metric. Equations \eqref{eq:feqgen}, \eqref{eq:vlasovgeneral}, and \eqref{eq:energymomentum} constitute the Einstein-Vlasov system in its general form. In order to fix the boundary conditions corresponding to an isolated system, we prescribe the spacetime to be asymptotically flat. In this form the system is very complex  from both an analytical and numerical viewpoint. We thus restrict the system to spherical symmetry which will be discussed in Section~\ref{sc:sphericalsym}. For a more elaborate discussion of the Einstein-Vlasov system we refer to~\cite{An2011,Rein95}. 

The Einstein-Vlasov system possesses a plethora of steady state solutions. For a stationary metric, the Killing vector  \mbox{$\partial/\partial t$} gives rise to the quantity  \mbox{$E = -g(\partial/\partial t,p^\alpha)$} which represents the particle energy  and is constant along geodesics. Hence, the ansatz
\begin{equation} \label{miceqstate}
f(x^a, p^b) = \varphi(E)
\end{equation}
satisfies the stationary Vlasov equation and reduces the system to the field equations. The ansatz function $\varphi$ can be used to generate stationary solutions which will be recalled in Section~\ref{ssc:steadystates}.

A physically motivated quantity of such a steady state is its (fractional) binding energy 
\begin{equation}\label{eq:bindeng}
	E_b = \frac{N-M}N,
\end{equation}
where $M$ is the ADM-mass and $N$ is the number of particles. For the definitions of $M$ and $N$ see Section~\ref{ssc:coordsys}.  Much of the literature concerning the stability for the Einstein-Vlasov system conjectures that stability behavior can be deduced from binding energy considerations leading to so-called binding energy hypotheses \cite{AbEtal1994, AnRe2006, Fack1970, Praktikum20, Ip1969_2,Ip1980,IT68, RaShTe1989_2, ShTe1985_2,Ze1971, Ze1966}.

Throughout the years, two binding energy hypotheses have often been intertwined and used equivalently. We distinguish between a weak and strong version in the following way: Let  \mbox{$\left (f_{z}\right )_{z> 0}$} be a family of asymptotically flat steady states to the Einstein-Vlasov system with finite mass and compact support parametrized by an appropriate redshift-factor $z$. \\

\noindent The \textit{weak binding energy hypothesis} holds if the steady states are stable at least up to the first local maximum of the binding energy curve parametrized by the redshift.  \\

\noindent The \textit{strong binding energy hypothesis} holds if the steady states are stable precisely up to the first local maximum of the binding energy curve parametrized by the redshift, and become unstable beyond this maximum.  \\ 

Furthermore, a related conjecture asserts that changes in stability behavior can only occur at extrema of the binding energy curve. Here stable and unstable always refers to stability for small, radial perturbations in the spherically symmetric case.  

The paper proceeds as follows: We first recall the historical development, results, and conjectures concerning the stability of steady states in the next section and present the ansatz functions used in the literature. We then review the polytropic case $k=1$ in more detail in Section~\ref{ssc:k=1}. In Section~\ref{sc:sphericalsym}, we introduce the Einstein-Vlasov system in the two coordinate systems---Schwarzschild coordinates and maximal areal coordinates---and recall the theory of steady states. The numerical method is sketched in Section~\ref{sc:numerics}. The equations of state which we use to obtain our results, are defined in Section~\ref{sc:ansatzfunction} along with interesting properties. As for the results of our work, we firstly provide strong evidence against the strong binding energy hypothesis in Section~\ref{ssc:evidencebinding} and secondly observe the new effect of multiple stability changes along a one-parameter sequence of steady states in Section~\ref{ssc:stabilitychanges}. We conclude this work with observations we deem relevant for the peculiar stability behavior found for our equations of state, and we state open questions for future work. 

\subsection{Historical context of the binding energy hypotheses}\label{ssc:history}
Historically, the first mention of binding energy as an indicator for the onset of instability goes back to Zel'dovich et al. In \cite{Ze1971,Ze1966} they simply assume that steady states beyond the first maximum of the binding energy are unstable. The authors claim that each successive extremum is associated with a new form of instability. Their argument mainly relies on considerations of the fluid case which are then adopted to the Vlasov matter case. 

The first work towards analytical results was initiated by Ipser and Thorne. In \cite{IT68} they derive a linearization of the Einstein-Vlasov system in Schwarzschild coordinates for general isotropic equations of state in hope of obtaining results for linear stability. From a conserved quantity of the linearized system a variational principle for the smallest eigenvalue is derived under the assumption that such eigenvalues exist. Furthermore, a criterion for the existence of a zero-frequency mode is given. They describe the strong binding energy hypothesis for isothermal clusters put forward by Zel'dovich as \enquote{\textit{highly unlikely}}. In the subsequent work~\cite{Ip1969_1} Ipser derived that collisionless star clusters are stable if the corresponding fluid star model is stable.

The first numerical results concerning stability were obtained in \cite{Ip1969_2}. One model considered is the so-called Maxwell-Boltzmann distribution which is defined by
\begin{equation}\label{eq:maxwellboltzmann}
	\varphi(E) = \begin{cases} K \exp \left ( -\frac{E}{T} \right ) , & E \leq E_0, \\
					0 , & E > E_0 ,
				\end{cases}
\end{equation}
with temperature $T$ measured by an observer at spatial infinity and a constant $K>0$. In \cite{Ip1969_2, Ze1966} it is argued that such a distribution is the eventual fate of a system including collisions. The relationship between the cut-off energy $E_0$ and the temperature $T$ is given by
\begin{equation}\label{eq:E0_T_dependence}
	E_0 = 1 -  \varepsilon T ,
\end{equation}	
where $\epsilon$ is constant. The steady state family is then parametrized by prescribing the central-to-surface redshift. Motivated by the fluid models, another set of distribution functions is constructed that fulfill a power-law relationship in the density $\rho$ and the pressure $p$. Some of these power-law models generate what is called an extreme core-halo configuration which makes them harder to handle numerically. Fackerell considered such a case in his related work \cite{Fack1970} and found inconclusive results on where instability sets in. 

In \cite{Ip1969_2} a variety of test functions were used in the variational principle mentioned above, in order to locate the point of change of stability. In the Maxwell-Boltzmann as well as in the power-law case the author finds that the steady states possess an exponentially growing mode for central redshifts  \mbox{$z_c \gtrapprox 0.5$} and are stable for  \mbox{$z_c \lessapprox 0.5$}. It is noted that the first binding energy maximum always appears in the same vicinity as  \mbox{$z_c \approx 0.5$} and that---for certain models---it should be possible to probe stability by looking at the behavior of the binding energy curve.

In 1980, it was shown by Ipser \cite{Ip1980} that isotropic clusters are linearly stable at least up to the first maximum of the binding energy curve. However, this proof lacks mathematical rigor since it assumes a turning point principle for the sign changes of the eigenvalues. Nevertheless, all the evidence considered up to today supports his claim. In Ipser's work it is also conjectured that stability changes must occur at extrema of the binding energy curve. This would lead to the expectation that general relativity can induce instabilities in nearly Newtonian steady states if local maxima appear arbitrarily close to $z_c = 0$.

In \cite{ShTe1985_1, ShTe1985_2} Shapiro and Teukolsky present their numerical method and results on the binding energy hypotheses in maximal isotropic coordinates. It is the first numerical simulation of the dynamics of the Einstein-Vlasov system concerning stability. The models for which stability is examined are the Maxwell-Boltzmann distribution \eqref{eq:maxwellboltzmann}, monoenergetic distributions defined by 
\[
	\varphi(E) = K \delta\left (1-\frac{E}{E_0}\right ),
\]
and explicit power-law distributions
\begin{equation}\label{eq:power-law}
	\varphi(E) = \begin{cases} 
				K \left (\frac{E}{E_0}\right )^{-2\delta} \left (1- \left (\frac{E}{E_0}\right )^{2} \right )^\delta, & E \leq E_0, \\
				0, & E > E_0 ,
				\end{cases}				
\end{equation}
where $\delta$ is constant. Across all of these models, steady states become unstable precisely at the maximum of the binding energy. They find that the latter does not coincide with  \mbox{$z_c \approx 0.5$} for  \mbox{$\delta = \frac{3}{2}$} and  \mbox{$\delta = \frac{5}{2}$} and, hence, discard that  \mbox{$z_c \approx 0.5$} is a signal for the onset of instability. 

Instead of using particle methods, a phase-space method is employed in \cite{RaShTe1989_2} to approximate solutions of the Einstein-Vlasov system. For clusters of the form \eqref{eq:power-law} the strong binding energy hypothesis is again confirmed. First studies into the axisymmetric Einstein-Vlasov system are conducted in \cite{AbEtal1994, ShTe1993}, where distribution functions similar to those above with an additional rotational component are used. The results indicate that a binding energy hypothesis might also hold in axisymmetry. 

If steady states are at least stable up to the first maximum of the binding energy, it is a natural question if one can construct a steady state family that has a monotonically increasing binding energy. Rasio et al.\ \cite{RaShTe1989_1} construct a family of steady states with reasonable physical properties for which the binding energy is monotonically increasing in $z_c$. In particular, the resulting steady states have finite mass and radius. They find that all investigated steady states are stable at least up to  \mbox{$z_c=3.75$} which pushes the predictability of the numerical method to its limits. They conclude that this family of steady states might be stable for all values of $z_c$. The results therefore support the binding energy hypotheses.

By considering \eqref{eq:maxwellboltzmann} as a two-dimensional family of steady states parametrized by $z_c$ and $\varepsilon$ from \eqref{eq:E0_T_dependence}, Bisnovatyi-Kogan et al.\ also find a family of steady states which have monotonically increasing binding energy in $z_c$, cf. \cite{Bisn1998}. However, stability is only deduced by assuming the validity of the weak binding energy hypothesis. 

The first investigation of anisotropic steady states was conducted by Andr\'{e}asson and Rein in \cite{AnRe2006} using maximal-areal coordinates. They consider polytropes of the form
\begin{equation}\label{eq:aniso_poly}
	\varphi(E,L) = \begin{cases} 
				 \left (1- \frac{E}{E_0}\right )^k (L-L_0)^l, & E < E_0 \text{ and } L > L_0, \\
				0, & \text{else},
				\end{cases}
\end{equation}
for  \mbox{$k\in\{0,1\}$, $l\in\{ 0,\frac12,\frac32\}$}, and  \mbox{$L_0 =\frac1{10}$} which lead to shell solutions with a vacuum region at the center. For collapse-promoting perturbations, the steady states collapse to a black hole exactly beyond the first binding energy maximum which thus supports the strong binding energy hypothesis. 

In a previous paper  \cite{Praktikum20} by the authors and colleagues, the results from \cite{AnRe2006} are confirmed and extended in multiple coordinate systems for different values of $k$, $l$, and $L_0$. In addition, isotropic polytropes, i.e.,
\begin{equation}\label{eq:iso_poly}
	\varphi(E) = \begin{cases} 
				\left (1- \frac{E}{E_0}\right )^k, & E < E_0, \\
				0, & E \geq E_0, 
				\end{cases}
\end{equation}
and the so-called king model
\begin{equation}\label{eq:king}
	\varphi(E) = \begin{cases} 
				\exp \left ( 1-\frac{E}{E_0}\right ) - 1, & E < E_0, \\
				0, & E \geq E_0, 
				\end{cases}
\end{equation}
are considered. Alongside finding evidence for the strong binding energy hypothesis again, the existence of heteroclinic orbits to new seemingly stable steady states is observed when dispersion-promoting perturbations are employed beyond the binding energy maximum. 

Recently, the mathematical study of stability for the linearized Einstein-Vlasov system was initiated by Had\v zi\'c and Rein in \cite{HaRe2013, HaRe2014}. The authors prove that steady states which are not too relativistic are linearly stable. More precisely, the positive coercivity of the second variation of the energy-Casimir functional is shown for small values of the central-to-surface redshift. Following that work, in \cite{HaLiRe2020} the strongly relativistic case is considered. It is proven that an exponentially growing mode for the linearized system exists when the central-to-surface redshift is large enough. Furthermore, an exponential trichotomy into a stable, unstable, and center space is shown. 
Note that this may not contradict the fact that steady states can be stable for arbitrarily large redshifts as reviewed above, because in \cite{HaLiRe2020} a fixed distribution function is prescribed, while in \cite{Bisn1998, RaShTe1989_1} the distribution function depends on multiple parameters. 

Overall, we conclude that all the numerical and---admittedly limited---analytical evidence prior to this work point towards the validity of the strong binding energy hypothesis. 

\subsection{Review of the polytropic case $k=1$}\label{ssc:k=1}
As a preparation for the investigation of more complex models in the upcoming sections, we briefly review the stability results from~\cite{Praktikum20} in the isotropic polytropic case $k=1$, i.e., a microscopic equation of state of the form \eqref{eq:iso_poly} with $k=1$.

To investigate the stability properties of stationary solutions we perturb them as weakly as possible and evolve the perturbed state. The specific perturbation scheme is described in Section~\ref{sc:results}.

\begin{figure}[h]
  \begin{subfigure}[h]{0.49\textwidth}
    \resizebox{!}{.72\textwidth}{\input{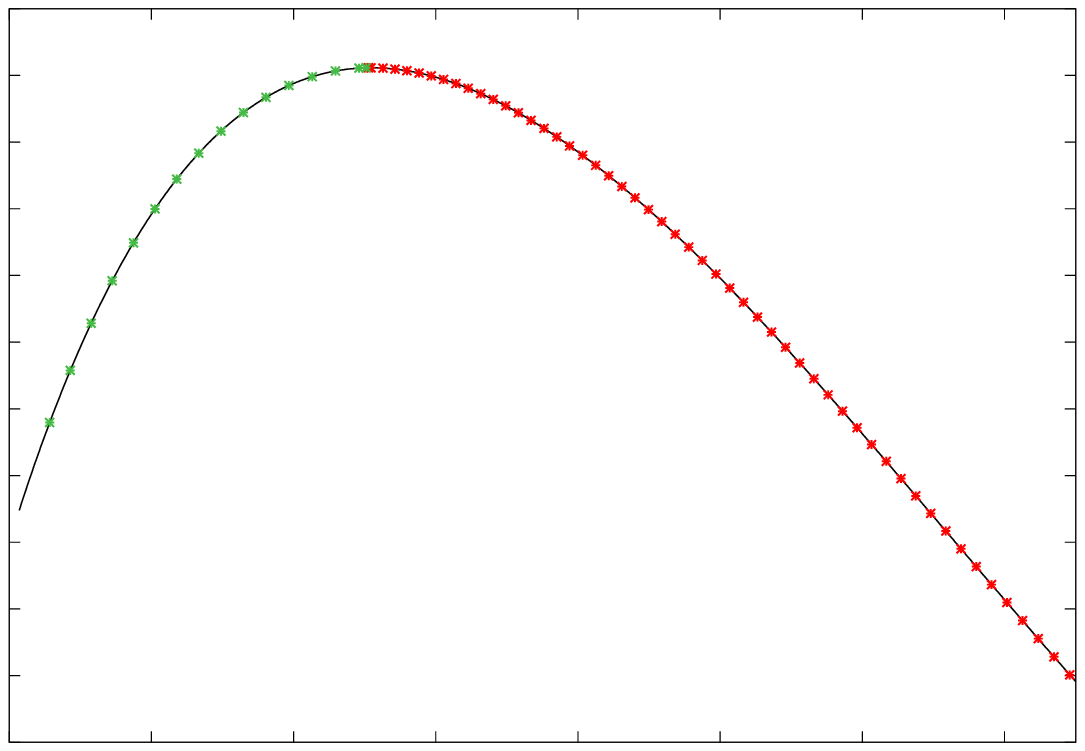}}\vspace*{-.16cm}
  \end{subfigure}
  \begin{subfigure}[h]{0.49\textwidth}
   \centering
    \resizebox{!}{.72\textwidth}{\input{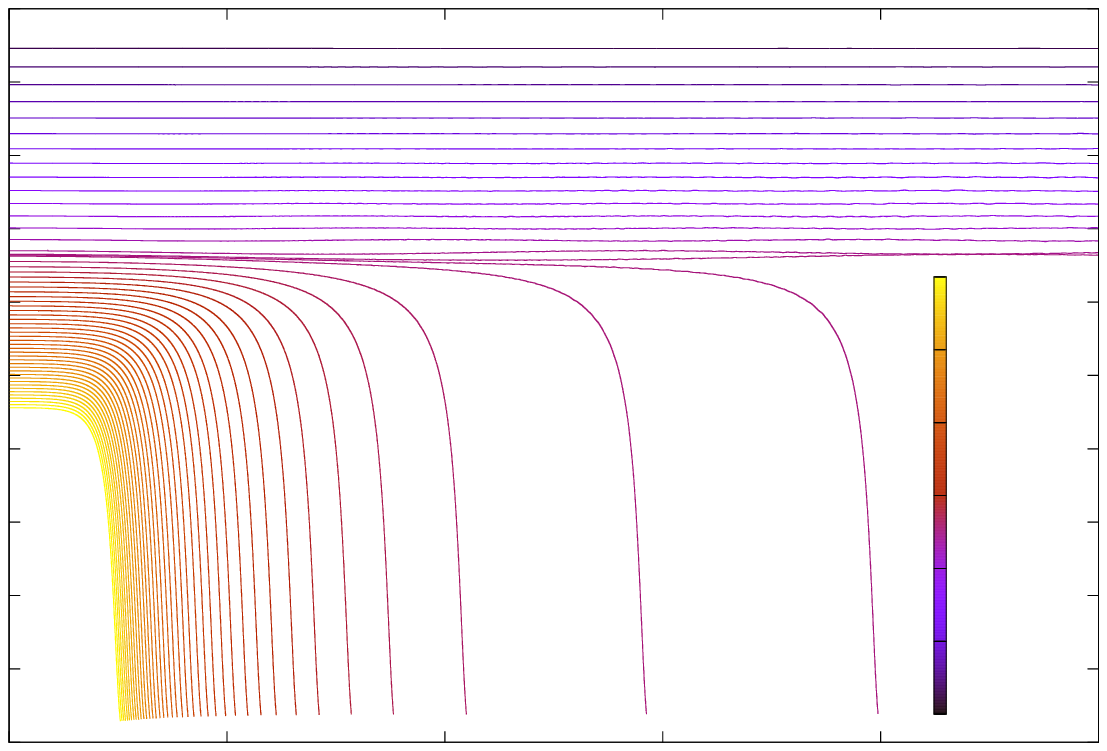}}\vspace*{-.16cm}
  \end{subfigure}
	\caption{The stability behavior of the isotropic case  \mbox{$k=1$} for different values of $z_c$ in maximal areal coordinates. On the left hand side the binding energy is plotted; green color corresponds to stable, red to unstable models. The right hand side shows the corresponding evolution of  \mbox{$\alpha(t,0)$}. 
	}
  \label{img:k=1_stab}
\end{figure}

The time evolution of several such perturbed states with different redshifts is depicted in Figure~\ref{img:k=1_stab} in the case of maximal areal coordinates. The plotted quantity  \mbox{$\alpha(t,0)$} is the value of the lapse function at the origin, see Section~\ref{ssc:coordsys} for a precise definition. It is a natural quantity to study since it is the relativistic counterpart of the gravitational potential at the spatial origin. Similar behaviors can be observed for other quantities as well as in other coordinates. For the definition of the parameter $z_c$ used to parametrize the different steady states instead of the redshift, we refer to Section~\ref{ssc:steadystates}.

%
%
For small values of $z_c$---corresponding to less relativistic settings---we observe a stable oscillation of the solution around the original equilibrium. For the spherically symmetric Vlasov-Poisson system, which is the non-relativistic limit of the Einstein-Vlasov system~\cite{RR92b}, similar oscillations have been observed numerically~\cite{RaRe2018} and have recently been proven to exist on the linear level for certain equilibria in~\cite{HaReSt21}.

Increasing $z_c$, we observe the onset of instability at the first maximum of the binding energy, cf.\ Figure~\ref{img:k=1_stab}. Thus, the strong binding energy hypothesis seems to hold true here. For the chosen type of perturbation, the instability manifests as collapse of the matter. From Figure~\ref{img:k=1_stab} it is also evident that the time after which a trapped surface forms is decreasing in $z_c$ within the unstable regime, i.e., slight perturbations of more relativistic models lead to a faster collapse; for a precise description see Section~\ref{sc:results}.

\section{The system in spherical symmetry}\label{sc:sphericalsym}
In order to cope with the complexity of the system, we consider the Einstein-Vlasov system only in the spherically symmetric, asymptotically flat case. Here spherical symmetry means $
	f(t,x,v)= f(t,Ax,Av) $ for $ A \in \mathrm{SO}(3).
$
Since general relativity allows for freedom in the choice of the coordinate system, we choose to analyze the system in Schwarzschild and in maximal areal coordinates. 
\subsection{The coordinate systems}\label{ssc:coordsys}
We shortly review the two coordinate systems. In the Schwarzschild case \cite{Rein95, RR92a} the line element takes the form 
\begin{align} 
  ds^2 = -e^{2\mu(t,r)} dt^2 + e^{2\lambda(t,r)} dr^2 + r^2 \left( d\theta^2 + \sin^2\theta d\psi^2 \right), \label{eq:metric}
\end{align}
where the metric coefficients  \mbox{$\mu=\mu(t,r)$},  \mbox{$\lambda=\lambda(t,r)$} depend only on the time  \mbox{$t\in\R$} and the radius  \mbox{$r\geq0$}.
In these coordinates, the time $t$ can be thought of as the proper time of an observer located at spatial infinity. For numerical and analytical reasons it is convenient to introduce Cartesian coordinates 
\[
x =(x^1, x^2, x^3)= r(\sin\theta \cos\psi, \sin \theta \sin \psi , \cos \theta ) \in \R^3
\]
and corresponding non-canonical momentum variables
\[
v^i = p^i + \left ( e^{\lambda} -1 \right ) \frac{x \cdot p}{r} \frac{x^i}{r} .
\]
Since we consider the asymptotically flat case and in order to guarantee a regular center, we impose the boundary conditions
\begin{align} \label{eq:boundary_conditions}
	\lim_{r \to \infty} \lambda \left( t, r \right) =\lim_{r \to \infty} \mu \left( t, r \right)= 0 = \lambda \left( t, 0 \right) , \quad t \in \R.
\end{align}
The Einstein field equations~\eqref{eq:feqgen} can be rewritten in terms of $\mu$ and $\lambda$. A sufficient set of equations in order to determine $\mu$ and $\lambda$ are 
\[
	e^{-2\lambda} \left( 2r \partial_r \lambda - 1 \right) +1 = 8 \pi r^2 \rho, \quad e^{-2\lambda} \left( 2r \partial_r \mu + 1 \right) -1 = 8 \pi r^2 p,
\]
where the source terms $\rho$ and $p$ are defined as 
\[
	\rho = e^{2\mu} \, T^{00}, \quad p = e^{-2\lambda} \, T^{ij}\frac{x_i x_j}{r^2}.
\]
The Vlasov equation takes the form
\begin{align}
	\partial_t f + e^{\mu - \lambda} \frac{v}{\sqrt{ 1 + \vert v\vert^2}} \cdot \nabla_x f - \left(\partial_t \lambda\, \frac{x\cdot v}r + \partial_r \mu\,e^{\mu - \lambda} \sqrt{ 1 + \vert v\vert^2} \right) \frac{x}{r} \cdot \nabla_v f = 0. \label{eq:EV_Vlasov} 
\end{align}
Here $|v|$ denotes the Euclidean length and $x\cdot v$ the Euclidean scalar product. The ADM mass 
\[
	M = \int_0^\infty 4 \pi r^2  \rho  \, dr 
\]
is conserved along solutions of the Einstein-Vlasov system. Other conserved quantities are all Casimir-functionals 
\begin{equation} \label{eq:casimir}
	\mathcal{C}(f) = \int_{\R^3}\int_{\R^3} e^{\lambda} \chi(f) \, dx \, dv ,
\end{equation}
where  \mbox{$\chi \in C^1(\R)$} with  \mbox{$\chi(0)=0$}. For the special choice  \mbox{$\chi(f) = f$} we get the number of particles $N$.

In the maximal areal coordinate system the structure is slightly different \cite{AnRe2006,Gue19,GueRe21}. Here the line element takes the form
\[
	ds^2 = (-\alpha^2 + a^2 \beta^2) dt^2 + 2a^2\beta dt dr + a^2 dr^2 + r^2( d\theta^2 + \sin^2\theta d\psi^2) ,
\]
where \mbox{ $\alpha=\alpha(t,r)$},  \mbox{$a=a(t,r)$}, and  \mbox{$\beta=\beta(t,r)$} are the metric coefficients. It is useful to define  \mbox{$\kappa =\frac{\beta}{r\alpha}$}. To fix the remaining coordinate freedom, we impose the maximal slicing gauge, i.e., each hypersurface of constant $t$ has vanishing mean curvature. Similarly to the Schwarzschild case, Cartesian coordinates $x$ and non-canonical momentum variables $v$ are introduced. In analogy to \eqref{eq:boundary_conditions}, the metric coefficients satisfy the boundary conditions
\begin{align} \label{eq:boundaryma} 
	 \lim_{r\to\infty} a(t,r) = \lim_{r\to\infty} \alpha(t,r) = 1 = a(t,0), \quad \beta(t,0) = 0.
\end{align}
The set of equations 
\begin{align*}
	 &\partial_r \left ( \frac{r}{2}\left (1-\frac{1}{a^2}\right ) \right ) = 4\pi\rho r^2 + \frac{3}{2} \kappa^2 r^2 , 
	\quad \partial_r \kappa = - 3\frac{\kappa}{r} -  4\pi a j,  \\
	&\partial_r \left (\frac{r^2\partial_r\alpha }{a} \right ) = (4\pi (\rho +S) + 6\kappa^2 ) r^2 a \alpha ,
\end{align*}
relates $\alpha$, $a$, and $\beta$ to $f$. These three equations are equivalent to \eqref{eq:feqgen} for the current metric. The source terms are given by the energy-momentum tensor through the relations
\[
	\rho = \alpha^{2} T^{00}, \quad j = a\alpha \left ( T^{0i} \frac{x_i}{r} + \beta T^{00} \right ), \quad S = \left ( \delta_{ij} + \left (a^{2} -1 \right ) \frac{x_i x_j}{r^2} \right ) T^{ij} .
\] 
The Vlasov equation becomes
\begin{align}\label{eq:vleqfull}
  	\partial_t f + \left ( \frac{\alpha}{a} \frac v {\sqrt{ 1 + \vert v\vert^2}} -\beta \frac xr \right )  \cdot \nabla_x f +
 	\left ( - \frac{\partial_r \alpha}{a} \sqrt{ 1 + \vert v\vert^2}  \,  \frac xr + \alpha \kappa \left ( v - 3 \frac{x\cdot v}r \frac xr \right ) \right ) \cdot \nabla_v f = 0   .
\end{align}
In maximal areal coordinates the ADM mass amounts to
\[
	M = \int_0^\infty (4\pi \rho + \frac 3 2 \kappa^2)r^2 \, dr,
\]
and the Casimir functionals can be defined as in \eqref{eq:casimir} with $a$ instead of $e^\lambda$ as the weight in the integral. 

The main difference between the two coordinate systems is the existence of a criterion for the formation of trapped surfaces. In Schwarzschild coordinates such a criterion does not exist, since these coordinates cannot cover an open region which contains a trapped surface. In maximal areal coordinates a trapped surface is present when
\begin{align}\label{eq:TS_cond_MA}
	\frac 1 {a(t,r)} - r \kappa(t,r) < 0 .
\end{align}
In this case, the expansion of both outgoing and ingoing null geodesics is negative at the time $t$ on the sphere of radius $r$. This signals the development of a spacetime singularity, cf.\ \cite{Pen1965}.

In Schwarzschild coordinates there exists a local existence and uniqueness result for smooth, compactly supported initial data together with a continuation criterion for such solutions, cf.\ \cite{Rein95,RR92a}. An analogous result holds in maximal areal coordinates, cf.\ \cite{Gue19,GueRe21}. The stability analysis in
\cite{HaLiRe2020, HaRe2013,HaRe2014} was carried out in Schwarzschild coordinates.

\subsection{Steady state solutions}\label{ssc:steadystates}

Despite the differences of the above-mentioned coordinate systems, for a given steady state solution $f_0$ in Schwarzschild coordinates it is easy to prove that $f_0$ is also a steady state in maximal areal coordinates and vice versa. The metric coefficients are related via \mbox{$\alpha_0 = e^{\mu_0}$, $a_0 = e^{\lambda_0}$}, and  \mbox{$\beta_0 = 0$}. For this reason, we discuss steady states only in Schwarzschild coordinates. 

The simplest way to obtain stationary solutions of the Einstein-Vlasov system is by writing the distribution function of the form
\begin{equation}\label{eq:general_ansatzEL}
	f_0 = \varphi(E,L)
\end{equation}
for suitable  \mbox{$\varphi\colon\R^2\to[0,\infty[$}, where  \mbox{$E = e^\mu \sqrt{1+|v|^2} $} is the particle energy and  \mbox{$L = |x \times v|^2$} can be interpreted as the square of the angular momentum. 
Observe that any sufficiently regular function of the form~\eqref{eq:general_ansatzEL} solves the Vlasov equation since $E$ and $L$ are integrals of motion. As in most of the physics literature mentioned above we only consider isotropic equilibria here, i.e., the case where 
\begin{equation} \label{eq:general_ansatz}
	f_0 = \varphi(E) = \Phi\left(1-\frac{E}{E_0} \right)
\end{equation}
is a function of the energy $E$ only. Here,  \mbox{$E_0\in]0,1[$} is the cut-off energy of the steady state, i.e.,  \mbox{$\varphi(E)=0$} for  \mbox{$E\geq E_0$} and $E_0$ is the minimal such value. The existence of $E_0$ is imposed in order to guarantee compact support and finite ADM mass. The ansatz function $\Phi$, which we also refer to as the microscopic equation of state, vanishes on  \mbox{$]-\infty,0]$}, and the only relevant part for the steady state are the values of $\Phi$ on  \mbox{$[0,1]$}. 

The existence of steady states of the form~\eqref{eq:general_ansatz} has, e.g.,\ been investigated in~\cite{RaRe2013}, and since we have to construct the steady states numerically we outline the arguments in the following. Mathematically we require  \mbox{$\Phi\colon\R\to[0,\infty[$} to be measurable and that there exist  \mbox{$\eta_0,c_1,c_2>0$} and  \mbox{$-\frac12<k<\frac32$} such that 
\[
c_1\,\eta^k \leq \Phi(\eta) \leq c_2\,\eta ^k, \quad \eta\in [0,\eta_0].
\] 
We exclusively consider the case $k=1$ in this work, cf. Section~\ref{sc:ansatzfunction}. Substituting the ansatz~\eqref{eq:general_ansatz} into the Einstein-Vlasov system reduces the time-independent system to an equation for $\mu_0$. Instead of solving the latter equation, it is more feasible to consider $y=\ln(E_0)-\mu_0$, which solves
\begin{equation}\label{eq:ssdgl}
	y' = - \frac{1}{1-\frac{8\pi}{r}\int_0^r  g(y(s))s^2 \, ds} \left( \frac{4\pi}{r^2} \int_0^r g(y(s)) s^2 \, ds + 4\pi r h(y(r)) \right).
\end{equation}
For the definitions of  \mbox{$g,h\in C^1(\R)$} we refer to~\cite{RaRe2013}. For every choice of \mbox{$y(0) = y_0 > 0$} a unique solution of~\eqref{eq:ssdgl} exists which yields a steady state with finite ADM mass and compact support~\cite{RaRe2013}; the cut-off energy is given by  \mbox{$E_0= e^{y(\infty)}$}. 
Hence, one given equation of state $\Phi$ yields a one-parameter family of steady states.  

Instead of $y_0$  we also use the redshift to parametrize the steady state family. The central redshift of a photon which is emitted at the center $r = 0$ and received at infinity is given by 
\begin{equation}\label{eq:centralredshift}
	z_c = \frac{e^{y_0}}{E_0}-1.
\end{equation}
In contrast, the central-to-surface redshift is given by  \mbox{$z_s = e^{y_0} -1$}. Note that the central redshift is not necessarily one-to-one with $y_0$ but in all of the following it is the case numerically. However, we still choose to use $z_c$ as it is prevalent in the physics literature.

\subsection{A scaling law}\label{ssc:scalinglaw}
For the numerical investigation it is helpful to know the behavior of solutions under a scaling transformation which has originally been derived in~\cite{Rein95}. 
We again only consider Schwarzschild coordinates, but emphasize that the analogous scaling law also applies for maximal areal coordinates.

Let  \mbox{$f: I\times \R^6 \rightarrow [0,\infty[$} be a solution of the Einstein-Vlasov system with metric coefficients $\lambda$ and $\mu$ on some time interval $I$ containing $0$.
Then for every  \mbox{$\gamma \in \R \setminus \{0\}$}, 
\begin{align}\label{eq:scaling}
	f_\gamma (t,x,v) = \gamma^2 f(\gamma t, \gamma x, v)
\end{align}
defines another solution of the Einstein-Vlasov system on  \mbox{$\gamma^{-1} I \times \R^6$}. Its metric coefficients  $\mu_\gamma$, $\lambda_\gamma$ are given by
\[
	\lambda_\gamma(t,r) = \lambda(\gamma t, \gamma r), \quad \mu_\gamma(t,r) = \mu(\gamma t, \gamma r),
\]
whereas the ADM mass $M$ and Casimir functionals $\mathcal C$ of the original solution $f$ are related to the respective quantities  \mbox{$M_\gamma$, $\mathcal C_\gamma$} of $f_\gamma$ via  \mbox{$M_\gamma = \gamma^{-1} M$} and  \mbox{$\mathcal{C}_\gamma = \gamma^{-1} \mathcal{C}$}. 

In particular, the scaling identity~\eqref{eq:scaling} can be used to rescale stationary solutions. Let 
\[
f_0(x,v) = \Phi\left(1-\frac{E(x,v)}{E_0}\right ) 
\]
be a time-independent solution of the Einstein-Vlasov system for some fixed equation of state $\Phi$ as specified in Section~\ref{ssc:steadystates}.
Then for every $K>0$ 
\[
	f_0^{(K)} (x,v) = K f_0(\sqrt K x,v)    
\]
defines another stationary solution. $f_0^{(K)}$ is in fact the unique steady state solution corresponding to the equation of state $K\Phi$ and the same value of $y_0$ as $f_0$. Observe in particular that~\eqref{eq:scaling} implies that the stability properties of $f_0$ and $f_0^{(K)}$ are equivalent and the binding energy curve is the same for all $K>0$. 


%
%

%
%
\section{The numerical method}\label{sc:numerics}
In order to study the stability of steady states of the Einstein-Vlasov system, we first have to compute a stationary solution as described in Section~\ref{ssc:steadystates}. This means we have to solve the integro-differential equation~\eqref{eq:ssdgl} with boundary condition  \mbox{$y(0)=y_0$} for a given equation of state $\Phi$ and parameter  \mbox{$y_0>0$}, which we do numerically by applying the midpoint method. Note that we only have to compute the steady state once, allowing us to use very high accuracy for this part. 

The actual time evolution of a perturbed steady state is simulated using a \emph{particle-in-cell scheme} which has also been used in \cite{AmAnRi21,AnRe2006,Praktikum20,RaRe2018,ReReSch}. It is known to converge in the case of Schwarzschild coordinates \cite{ReRo} and also for the spherically symmetric Vlasov-Poisson system \cite{Sch}. The numerical method is explained in more detail in \cite{AnRe2006, Praktikum20,ReRo}, which is why we only sketch it here.

The idea of a particle-in-cell scheme is to split the support of the initial distribution function into distinct cells and place a numerical particle into each cell.
Each numerical particle represents the contribution of its cell, i.e., it carries a weight depending on the distribution function and the cell size.
To place these particles in phase space, we use variables adapted to spherical symmetry in order to reduce the dimension of phase space from $6$ to $3$. More precisely, we set up a grid of equidistant points in the spatial direction \mbox{ $r=\vert x\vert$} and place a cluster of particles with different momenta at every radial grid point. The amount of numerical particles placed at each radial grid point depends on the structure of the initial state. For instance, we have to place more particles within a dense region in order to represent such distributions properly, especially in the case of a core-halo structure, cf.\ Section~\ref{sc:ansatzfunction}.

The numerical particles are propagated according to the characteristic system corresponding to the Vlasov equation \eqref{eq:EV_Vlasov} and \eqref{eq:vleqfull} respectively. We use the classical fourth order Runge-Kutta (RK4) method for solving the resulting ODE. We emphasize that the particle propagation process is performed in Cartesian coordinates in order to prevent errors near the spatial origin. After each time step, the metric components are updated using RK4 based on the new positions of the numerical particles; the momentum integrals in the matter quantities are computed by summing up the contributions of each cell.

To investigate the stability behavior of a given steady state, we have to perturb it slightly. We apply the same dynamically accessible perturbation scheme as in~\cite{Praktikum20}, i.e., during an initial time interval  \mbox{$[0,T_{pert}]$} we add the divergence free term  \mbox{$(0,\gamma\frac xr)$} for prescribed  \mbox{$\gamma\approx0$} to the right-hand side of the characteristic system used in the particle propagation. Observe that any physically viable perturbation, e.g., by some external force, should preserve all Casimir functionals \eqref{eq:casimir}, and the aforementioned perturbations particularly have this feature. We emphasize however that we have tested several other perturbation schemes as well, but there is no indication that the phenomena described in the upcoming sections depend on the specific type of the perturbation.

For the simulations performed in this work we use  \mbox{$T_{pert} \approx 10M$}. To control the intensity of the perturbation, we further prescribe $\epsilon_{pert}$ of the order $10^{-4}$ and choose $\gamma$ such that the relative error of $e^{\mu(t,0)}$ and  \mbox{$\alpha(t,0)$} respectively between $t=0$ and $t=T_{pert}$ is close to $\epsilon_{pert}$. The same procedure to determine $\gamma$ has also been used in \cite{Praktikum20} and leads to an absolute value of $\gamma$ between roughly $10^{-3}$ and $10^{-5}$. As described in \cite{Praktikum20}, the sign of $\gamma$ decides whether the solution is nudged towards dispersion ($\gamma>0$) or towards collapse ($\gamma<0$).
Furthermore, we choose the radial and time step size to be in the order of magnitude of $10^{-3}$.
The total number of numerical particles is between $10^7$ and $10^8$. 

To monitor the reliability of our simulations, we track the time evolution of $M$ and $N$---which are analytically conserved---among other quantities. The above set of parameters leads to relative errors of the former quantities of the order $10^{-6}$ after  \mbox{$t=1000M$}, even in the case of a collapse, which is a notable improvement compared to prior investigations \cite{Praktikum20}.

In order for the simulations to handle such large amounts of particles within a reasonable time-frame, our programs are parallelized using the Pthreads API in \texttt{C++}. Note that parallel computing fits very well with the particle-in-cell scheme \cite{KoRaRe2013}.

\section{The equations of state}\label{sc:ansatzfunction}
We now define the equations of state $\Phi$ which we use in this work. As reviewed in Section~\ref{ssc:history}, the binding energy hypotheses were only tested and confirmed for a limited range of equilibrium models, even though $\Phi$ can be chosen quite generally. This lack of diversity motivates the use of different functions. In the analytical work on stability in the isotropic case it is always required that  \mbox{$\Phi'>0$}, i.e.,  \mbox{$\varphi'<0$} on the support of the solution. This property is reasonable physically, since it implies that there are more low-energy than high-energy stars, cf. \cite{IT68}. All the equations of state considered here particularly satisfy this monotonicity assumption. Strictly speaking, the ansatz functions are not differentiable at one point, but our results are not affected by this. 

We use a whole family of ansatz functions which are piecewise linear. More precisely, $\Phi_i$ is defined by
\begin{equation}\label{eq:n_family}
	 \Phi_i(\eta) \coloneqq \begin{cases}
    \frac{\eta}{10}, & \text{if } 0 < \eta \leq \frac i{1000}, \\
  	\frac{i-100000}{10i-10000} \eta + \frac{99i}{10i-10000}, & \text{if }  \frac i{1000}< \eta, \\
  	0, & \text{else,}
  \end{cases}
\end{equation}
for  \mbox{$0\leq i \leq 1000$}. Observe that for  \mbox{$\eta < \frac i{1000}$}, $\Phi_i(\eta)$ equals the polytropic ansatz  $\eta_+^k$ with  \mbox{$k=1$} up to a factor. From  \mbox{$\frac i{1000}$} onwards the function is simply continued by the unique straight line that connects the first part of the function continuously with the fixed value  \mbox{$\Phi_i(1)=10$}. 

Note that we do not suggest that they are physically relevant models arising in nature. The main motivation for choosing these models lies in their new features and mathematical simplicity. 


\begin{figure}[h]
	\begin{center}
   	\centering
   		 \resizebox{!}{.52\textwidth}{\input{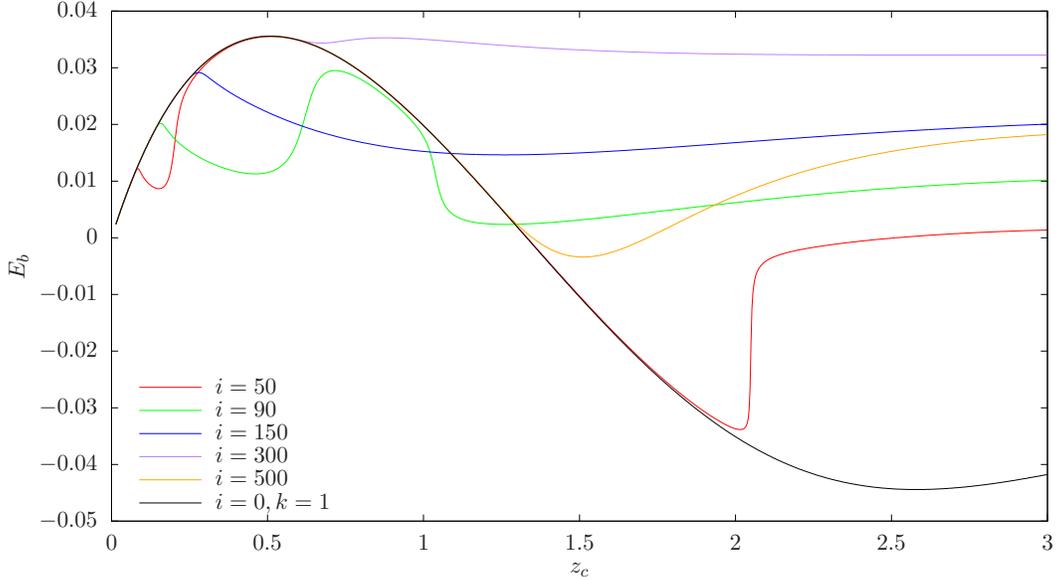}}\vspace*{-.16cm}
    \end{center}
  \vspace*{-.5cm}
	\caption{The binding energy curves for different models. We obtain a local maximum near $z_c=0$ for $i=50$ and two pronounced maxima for  \mbox{$i=90$}. The second maximum vanishes for  \mbox{$i=150$}. The maximum then drifts upwards to the right while the binding energy stays very much positive for  \mbox{$i=300$}. The binding energy eventually dips below zero again for  \mbox{$i=500$}. In the limit  \mbox{$i=1000$} we obtain the same curve as for  \mbox{$k=1$} and  \mbox{$i=0$}.   
	}
  \label{img:ansatzfunction}
\end{figure}

\begin{figure}[h]
	\begin{center}
	\centering
		 \resizebox{!}{.53\textwidth}{\input{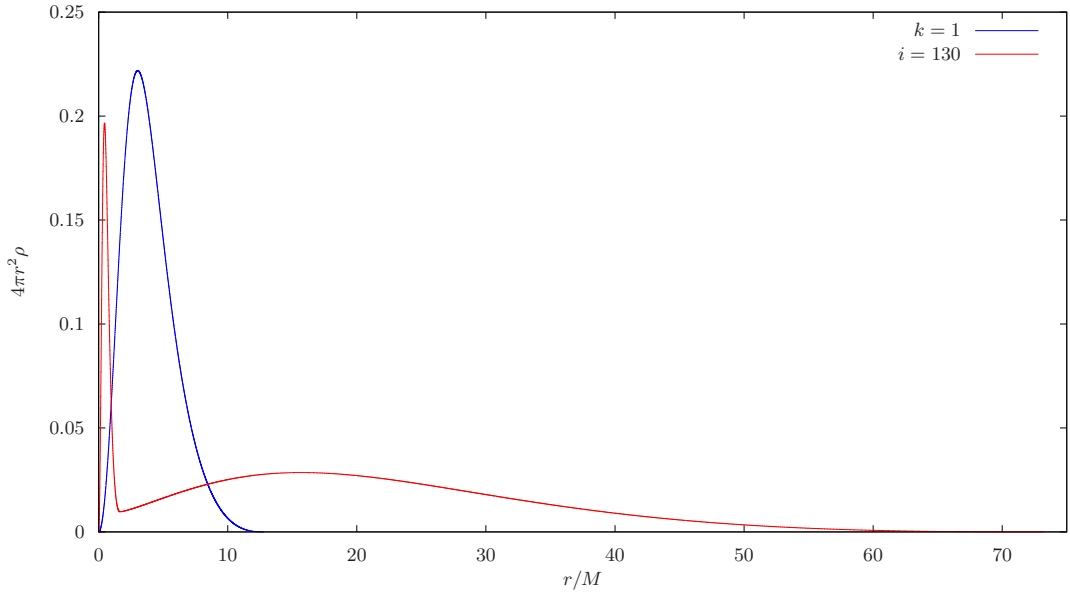}}
	\end{center}
	\vspace*{-.5cm}\caption{The mass density of a stable core-halo configuration for  \mbox{$i=130$} compared with the unstable polytropic case  \mbox{$k=1$} for the same value of  \mbox{$z_c = 0.57$}. The total mass is rescaled to unity for both models.
	} 
	\label{img:core_halo}
\end{figure}

Loosely speaking, the steep incline for  \mbox{$\eta > \frac i{1000}$} means that low-energy particles are assigned a high weight in the distribution function. Since the low-energy particles can be found only near the origin, the core of the steady state gets extremely dense for larger values of $z_c$. Around that dense core a long tail of particles with low weights in the distribution function forms. In the literature this long tail is sometimes called a Newtonian halo, cf.\ \cite{Bisn1998, Fack1970}. We refer to such steady states as core-halo configurations. An exemplary mass density for a core-halo configuration is plotted in Figure~\ref{img:core_halo}.

The main interest, of course, lies in the behavior of the binding energy curves. A rich variety of features arises for various values of $i$. We depict a sample of binding energy curves in Figure~\ref{img:ansatzfunction} which represent the most important properties. 

For $i=0$ the equation of state is equivalent to the polytropic case $k=1$ covered in Section~\ref{ssc:k=1}, since it can be obtained by setting  \mbox{$K=\frac1{10}$}, cf.\ Section~\ref{ssc:scalinglaw}.

For small values of $i$, e.g.,  \mbox{$i=50$}, a new local maximum in the binding energy curve appears continuously from the origin  \mbox{$z_c=0$}. Numerically, it appears as though the first local maximum can be generated arbitrarily close to  \mbox{$z_c=0$}. 

By increasing $i$, e.g., to  \mbox{$i=90$},  we obtain two pronounced local maxima for  \mbox{$z_c \lesssim 0.7$}, i.e., in the domain where stability changes usually appear. For even larger values of $i$, e.g., \mbox{$i=150$}, the second local maximum vanishes while the first local maximum continuously drifts up to the right in the \mbox{$(z_c, E_b)$} diagram. Furthermore, the binding energy becomes strictly positive and almost constant after the first binding energy maximum. 

When $i$ increases further, the binding energy for large values of $z_c$ decreases again. Eventually, for  \mbox{$i=1000$}, we again obtain the original polytropic case \mbox{$k=1$} since \mbox{$\Phi_0 = 100 \, \Phi_{1000}$}. 

In conclusion, the following questions arise: 
 Does instability always set in at the first local maximum of the binding energy, even if said maximum is very close to  \mbox{$z_c =0$}? If not, where does instability set in? Can stability change multiple times, i.e., from stable to unstable and back to stable? Does the presence of two pronounced local maxima have an effect?

We will try to answer these questions in the next two sections. Another interesting observation is the strict positivity of the binding energy for certain values of $i$. However, the implication on the existence of fully dispersing solutions is beyond the scope of this paper.

%
%


%
%
\section{Results}\label{sc:results}

We now investigate the stability properties of the equilibria given by the equations of state introduced in the previous section. For selected values of $i$ we consider the steady states induced by $\Phi_i$ and several different values of $z_c$. To uncover the stability properties of such stationary solutions, we perturb them using the dynamically accessible perturbation scheme introduced in Section~\ref{sc:numerics} and evolve the perturbed state over time. We restrict the discussion to collapse-promoting perturbations, since the transition from stable to unstable equilibria is most evident for this type of perturbation; corresponding to $\gamma < 0$ in our case. This is due to the fact that for dispersion-promoting perturbations, the dispersing behavior sets in continuously and is therefore hard to detect close to the stable regime; see~\cite{Praktikum20} for a detailed discussion. 

We emphasize, however, that we have perturbed several steady states which are stable against the aforementioned collapse-promoting perturbation using numerous other perturbations---also including non-dynamically accessible ones---but never observed any indication of unstable behavior.
\begin{figure}[H]
	\input{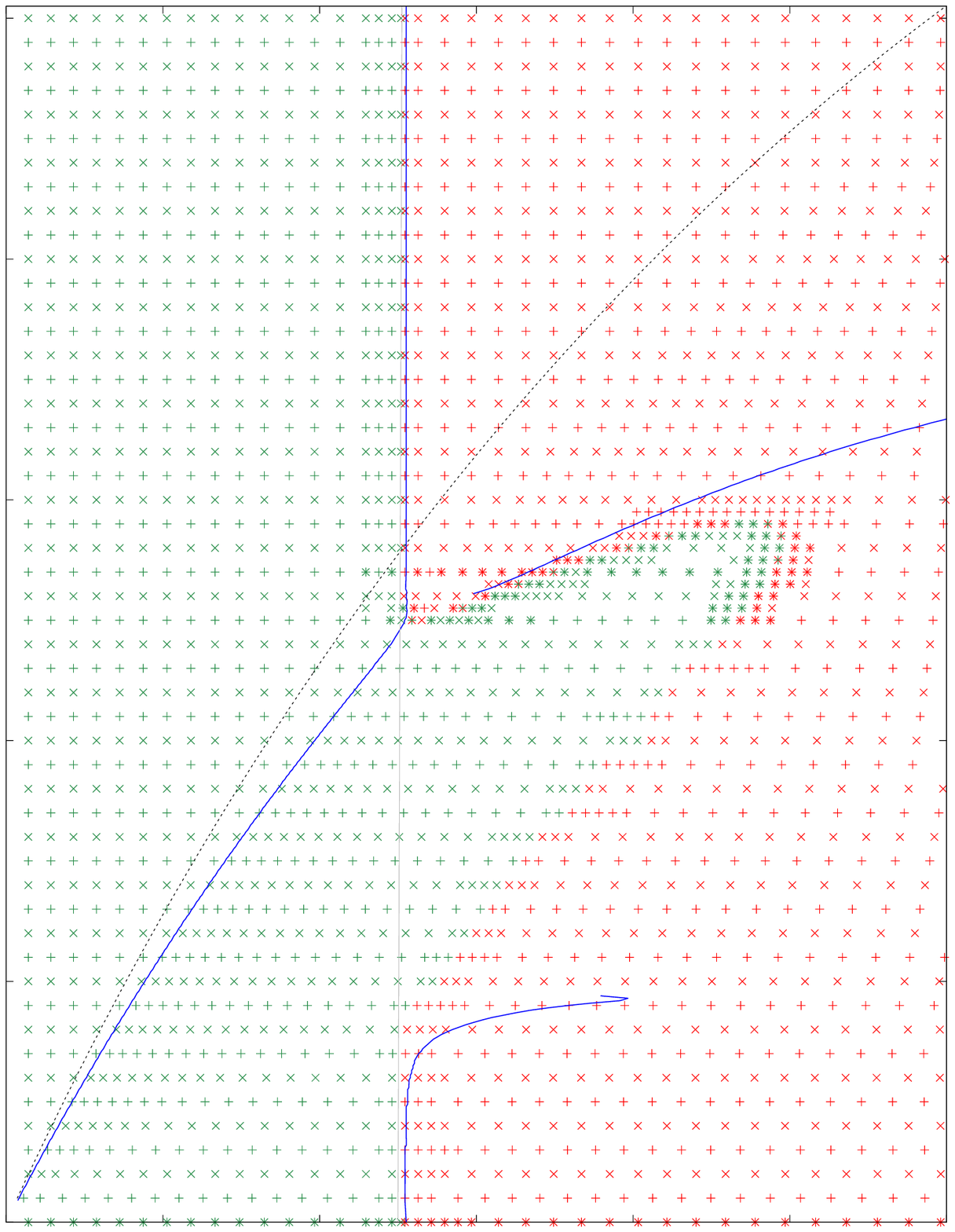}
	\vspace*{-.4cm}\caption{Stability of the equations of state $\Phi_i$ for different $z_c$. Green color indicates stable, red color unstable steady states in Schwarzschild coordinates ($+$) or in maximal areal coordinates ($\times$). Blue lines represent local binding energy maxima. The area above the dashed line corresponds to the polytropic case $k=1$. The gray line represents  \mbox{$z_c=\frac12$}. 
	}
	\label{img:aal}
\end{figure}
We thus refer to a stationary solution as stable if it does not collapse until a prescribed time $T$ after being perturbed as described above; otherwise we call it unstable. In the following we use \mbox{$T=1000 M$} but note that usually the collapse happens much faster. To verify whether a solution has collapsed, we check for the criterion~\eqref{eq:TS_cond_MA} in maximal areal coordinates. In Schwarzschild coordinates such an analytical criterion for a collapse does not exist, but we define a solution to be collapsed if the value of $e^\mu$ becomes sufficiently small at the spatial origin.

For each equation of state $\Phi_i$ under consideration, we limit the analysis to  \mbox{$z_c \leq 2$}---corresponding to \mbox{$y_0 \lesssim 1$}---since the solutions are harder to handle for larger $z_c$ from a numerics point of view, especially in the case of core-halo configurations as illustrated in Figure~\ref{img:core_halo}. The results of this extensive stability analysis are presented in Figure~\ref{img:aal}. Note that we only plot $z_c \leq 1.2$ as all models with higher values of $z_c$ are unstable.

As explained in Section~\ref{sc:ansatzfunction}, $\Phi_0$ and $\Phi_{1000}$ both correspond to a rescaling of the isotropic polytropic case  \mbox{$k=1$} discussed in Section~\ref{ssc:k=1}. This causes the stability properties at the top and at the bottom of the figure to be equal, since rescaling does not affect stability, cf.\ Section~\ref{ssc:scalinglaw}.

However, the attentive reader will have noticed that Figure~\ref{img:aal} is limited to  \mbox{$0\leq i\leq500$}. The reason for this is the following:\ During the steady state computation, the ansatz function $\Phi_i$ is only evaluated in the range  \mbox{$[0,1-e^{-y_0}]$}. The resulting stationary solution is thus identical to a rescaled version of the isotropic polytrope  \mbox{$k=1$} for small values of $y_0$ or $z_c$; the threshold $z_c$-value for this identity is given by the dashed line in Figure~\ref{img:aal}. Increasing $i$ enlarges the $z_c$-range where $\Phi_i$ is equivalent to the polytrope  \mbox{$k=1$}, and for  \mbox{$i\approx500$} this range covers the whole fixed $z_c$-range under consideration.

Furthermore, observe that Figure~\ref{img:aal} covers the results in Schwarzschild coordinates \mbox{(\enquote{$+$}-symbol)} as well as in maximal areal coordinates (\enquote{$\times$}-symbol), and the results in both coordinate systems are remarkably coherent. This is by no means obvious, since a coordinate invariant theory for stability does not exist in general relativity. 

Besides the stability behavior of the aforementioned steady states, Figure~\ref{img:aal} also shows the locations of the first two maximizers of the binding energy along the steady state family for one fixed equation of state as well as the redshift $z_c=\frac12$. As discussed in Section~\ref{ssc:history}, all these points have been conjectured to be linked to the stability behavior in the past. Whether or not this is the case in the present investigation will be discussed in the following sections.

\subsection{Evidence against the strong binding energy hypothesis}\label{ssc:evidencebinding}
As recalled in Section~\ref{ssc:history}, all numerical evidence prior to this work has confirmed the strong binding energy hypothesis, i.e., the onset of instability occurs precisely at the first maximum of the binding energy curve. However, our results clearly show that this hypothesis is not true for general isotropic equations of state. 

From Figure~\ref{img:aal} we conclude that the stability changes close to the second binding energy maximum for small values of $i$. This is consistent with the limit  \mbox{$i \to 0$} and thus the polytropic case $k=1$, cf.\ Section~\ref{ssc:k=1}. Note that the development of a local maximum for small $z_c$ and the continuous behavior of the onset of instability already bodes ill for the strong binding energy hypothesis. 

\begin{figure}[h]
  \begin{subfigure}[h]{0.49\textwidth}
    \resizebox{!}{.72\textwidth}{\input{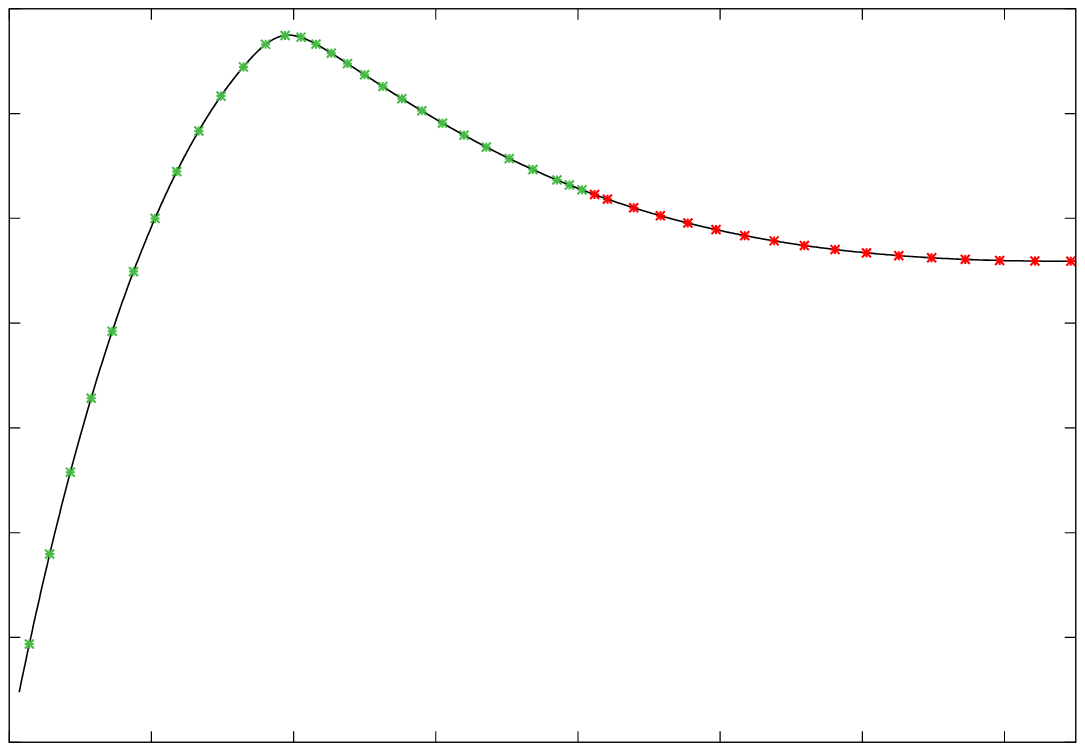}}\vspace*{-.16cm}
  \end{subfigure}
  \begin{subfigure}[h]{0.49\textwidth}
   \centering
    \resizebox{!}{.72\textwidth}{\input{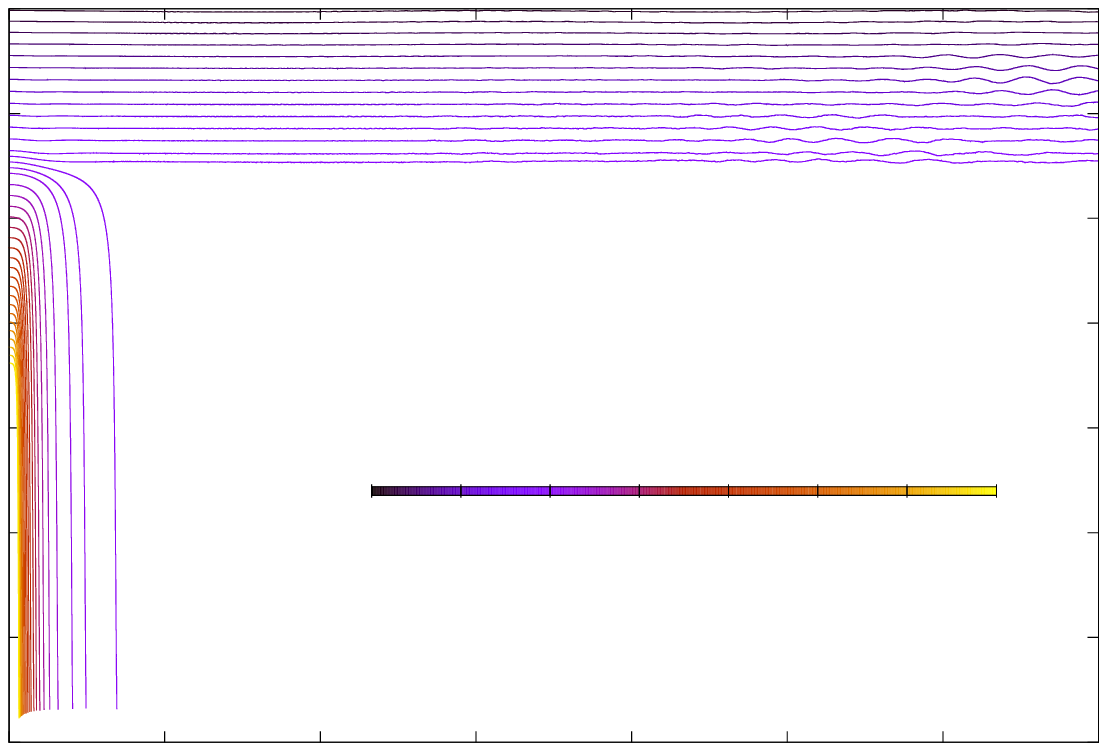}}\vspace*{-.16cm}
  \end{subfigure}
	\caption{The stability behavior of the equation of state $\Phi_{200}$ for different values of $z_c$ in maximal areal coordinates.  On the left hand side, the binding energy is plotted; green color corresponds to stable, red to unstable models. The right hand side shows the corresponding evolution of $\alpha(t,0)$.
	}
  \label{img:i=200_stab}
\end{figure}

As described in Section~\ref{sc:ansatzfunction}, the second maximum disappears when increasing $i$. Therefore, the change of stability cannot be deduced from binding energy considerations for \mbox{$ 100 \leq i \leq 250$}. For the purpose of illustration we will discuss this phenomenon for $\Phi_{200}$ explicitly: Figure~\ref{img:i=200_stab} depicts the stability behavior as well as the binding energy curve in this case. The steady states are stable for  \mbox{$z_c \leq 0.787 $} and unstable for  \mbox{$z_c \geq 0.806$} despite the first binding energy maximum being attained  at  \mbox{$z_c \approx 0.394$}. In addition, the change in stability cannot be attributed to other local extrema of the binding energy curve.

Increasing $i$ even further we observe that the onset of instability again coincides with the first binding energy maximum as the domain where the ansatz function is equivalent to $k=1$ increases, as can be seen from the dashed line in Figure~\ref{img:aal}. 

Remarkably, our results indicate that the weak binding energy hypothesis holds true across all models we considered with possible exceptions of borderline cases very close to the maximum. Note that it is impossible to determine the exact point where stability changes in a numerical investigation due to computational errors; this is why we do not consider such borderline cases as evidence against the weak binding energy hypothesis.     

%
%
\subsection{Multiple stability changes for a fixed equation of state}\label{ssc:stabilitychanges}
\begin{figure}[H]
  \begin{subfigure}[h]{0.49\textwidth}
    \resizebox{!}{.72\textwidth}{\input{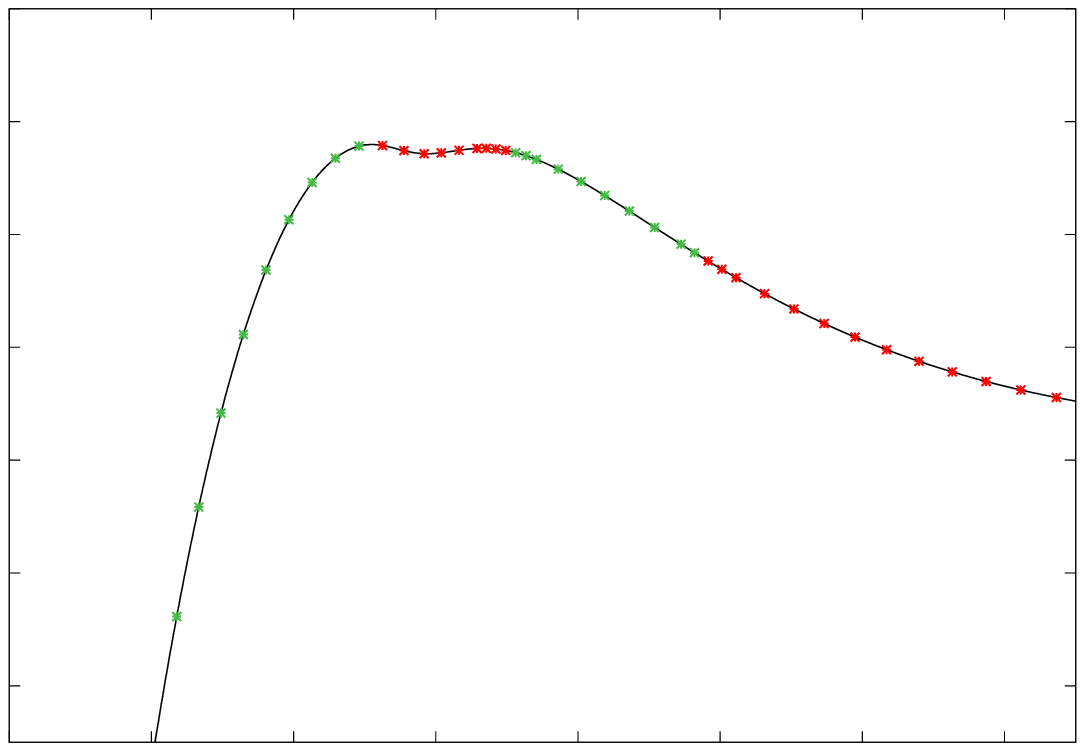}}\vspace*{-.16cm}
  \end{subfigure}
  \begin{subfigure}[h]{0.49\textwidth}
   \centering
    \resizebox{!}{.72\textwidth}{\input{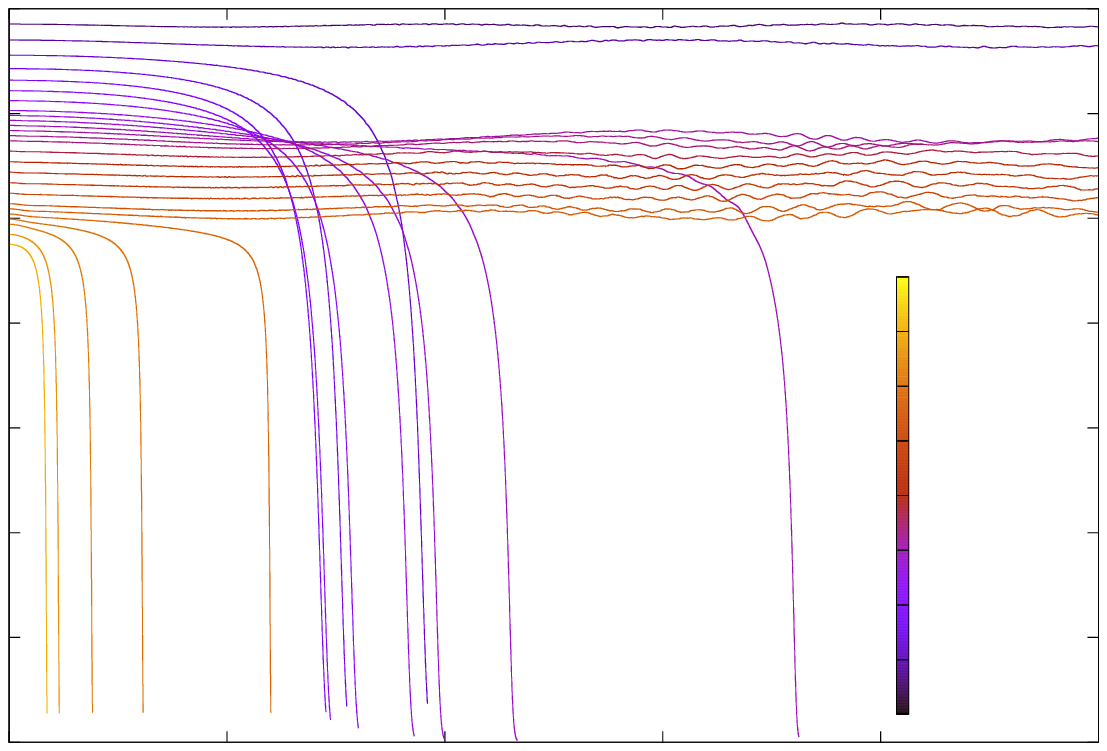}}\vspace*{-.16cm}
  \end{subfigure}
	\caption{The stability behavior of the equation of state $\Phi_{270}$ for different values of $z_c$ in maximal areal coordinates.  On the left hand side, the binding energy is plotted; green color corresponds to stable, red to unstable models. The right hand side shows the corresponding evolution of  \mbox{$\alpha(t,0)$}. 
	}
  \label{img:i=270_stab}
\end{figure}

We now discuss another exceptional behavior visible in Figure~\ref{img:aal}. For equations of state $\Phi_i$ with \mbox{$255 \leq i \leq  290 $} the stability changes twice. More precisely, there exist stable steady states with larger $z_c$ than unstable ones with the same equation of state. In the case of \mbox{$i=270$} this peculiar behavior is shown in Figure~\ref{img:i=270_stab}. In maximal areal coordinates the steady states are stable for \mbox{$z_c \leq 0.492 $} after which they are initially unstable. In fact, the first change in stability occurs very close to the first binding energy maximum. Increasing $z_c$ we observe another domain of stable steady states. This second stability domain corresponds to \mbox{$ 0.712 \leq z_c \leq 0.964$}; note that these steady states are stable for much longer than depicted in Figure~\ref{img:i=270_stab}. The lower bound of this range is located near the second binding energy maximum. We also observe that the binding energy curve is rather flat for the crucial values of $z_c$. For \mbox{$0.983 < z_c$} instability sets in again after which we do not see any evidence for further stability changes.  

\begin{figure}[H]
  \begin{subfigure}[h]{0.49\textwidth}
    \resizebox{!}{.72\textwidth}{\input{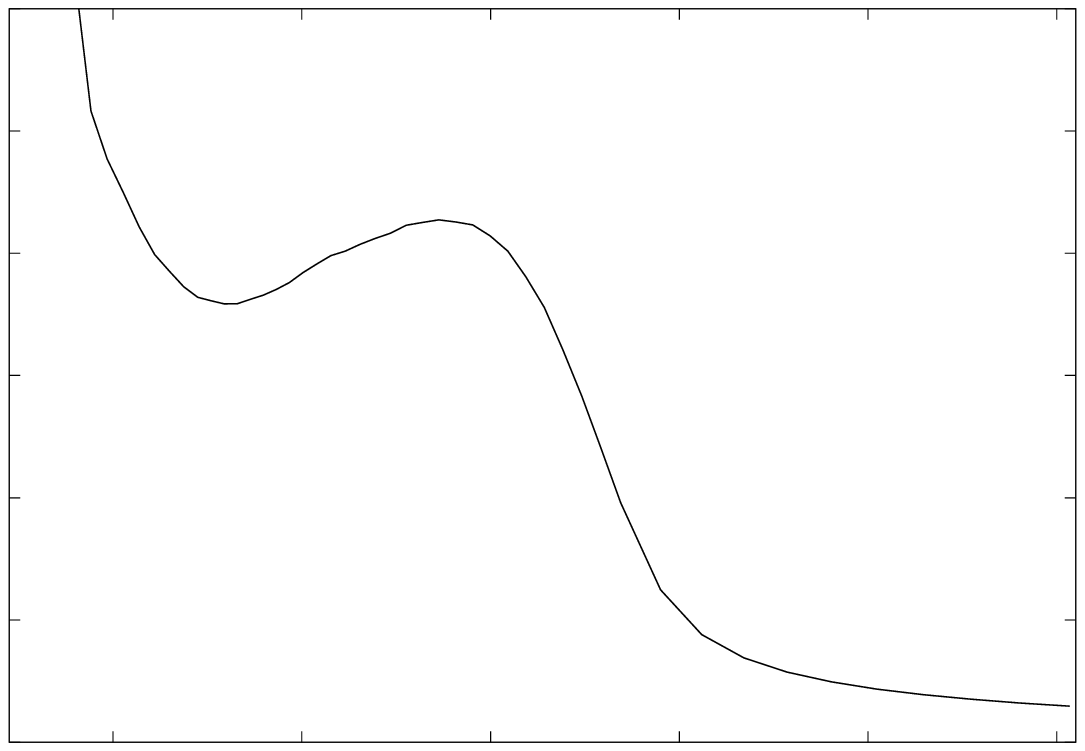}}\vspace*{-.16cm}
      \end{subfigure}
  \begin{subfigure}[h]{0.49\textwidth}
   \centering
    \resizebox{!}{.72\textwidth}{\input{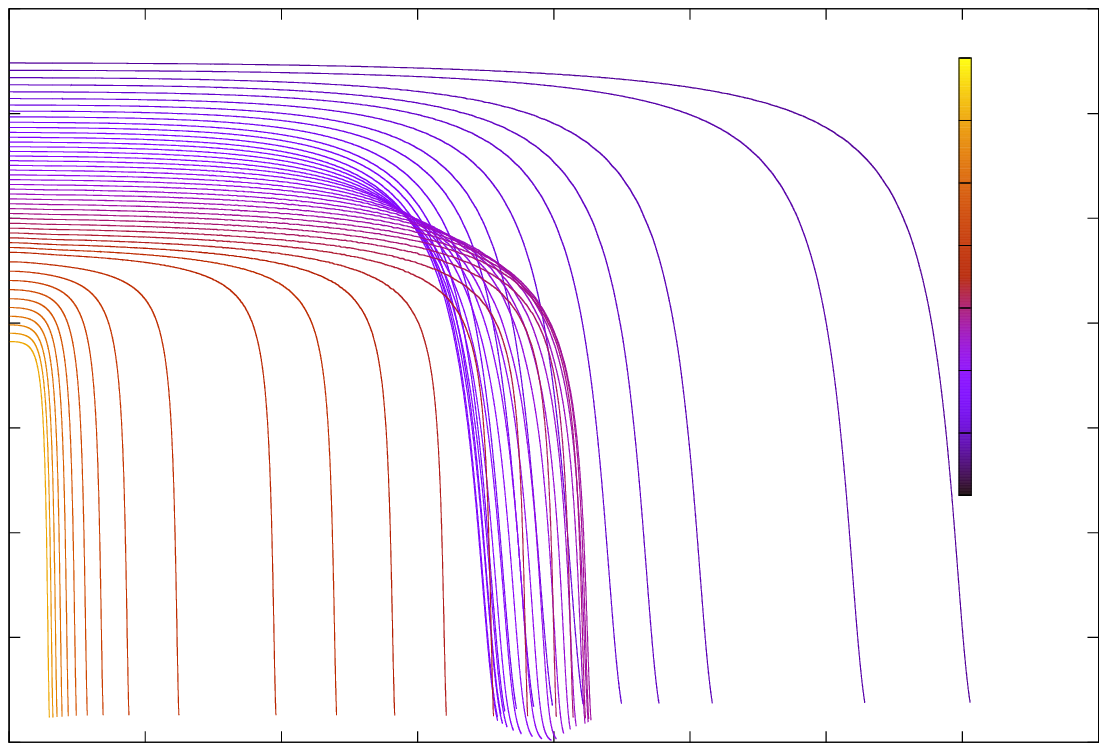}}\vspace*{-.16cm}
  \end{subfigure}
	\caption{Behavior of unstable solutions of the equation of state $\Phi_{310}$ for different values of $z_c$ in maximal areal coordinates. On the left hand side, the collapse time $t_{TS}/M$ is plotted. The right hand side shows the corresponding development of  \mbox{$\alpha(t,0)$}. 
	}
  \label{img:i=310_stab}
\end{figure}

Even though no multiple stability changes exist for \mbox{$i \geq 295$}, 
a remnant of the second stability domain can still be observed. Recall that the collapse time $t_{TS}$, i.e., the first point in time at which a trapped surface is formed \eqref{eq:TS_cond_MA}, is monotonically decreasing in $z_c$ for the polytropic case \mbox{$k=1$} as seen in Section~\ref{ssc:k=1}. In contrast, the collapse time generally increases the closer the steady state is located to the stable regime and thus the collapse time is not necessarily monotonically decreasing for one fixed equation of state. This effect is illustrated in Figure~\ref{img:i=310_stab} for \mbox{$i=310$}. Here the larger collapse time at \mbox{$z_c\approx 0.95$ } seems to be connected to the existence of the second stability domain for smaller values of $i$.  

It should be noted that steady states with higher values of $z_c$ require more careful numerical treatment, since, e.g., the value of the energy-density at the origin $\rho(0)$ enlarges rapidly. In order to ensure that the stability behavior is of physical rather than numerical nature, we have checked the cases where stability changes with higher numerical accuracy, but did not observe other qualitative behavior. We are thus convinced that our results represent the mathematical stability behavior to high accuracy. 

\begin{figure}[h]
	\begin{center}
	\centering
		 \resizebox{!}{.53\textwidth}{\input{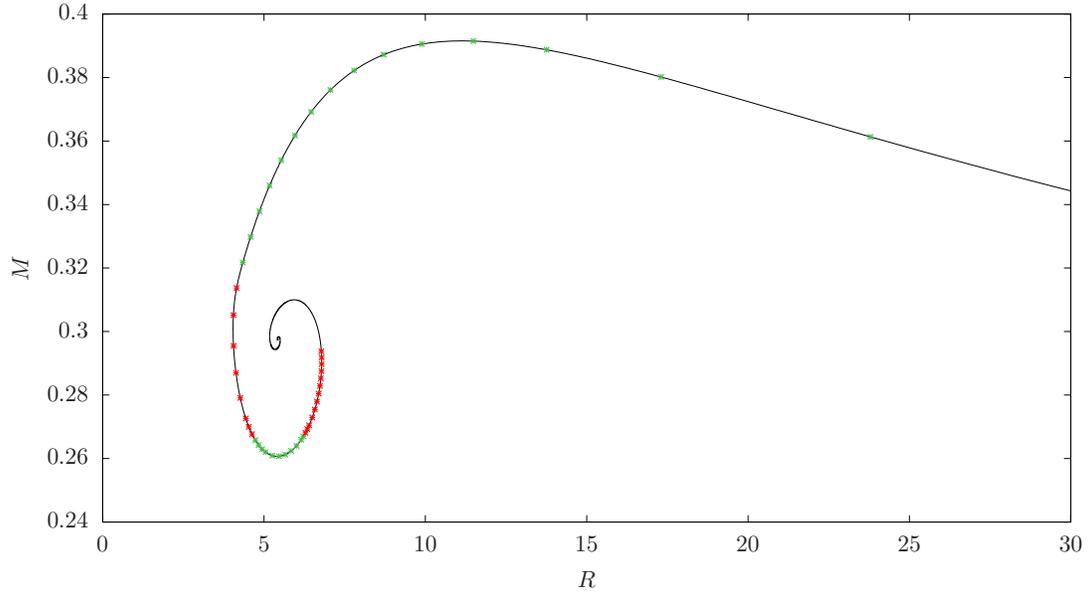}}
	\end{center}
	\vspace*{-.5cm}\caption{Mass-radius curve for the equation of state $\Phi_{270}$. As $z_c$ increases $(R,M)$ moves into the spiral.
		The color shows the stability behavior in Schwarzschild coordinates; green corresponds to stable, red to unstable models.} 
	\label{img:massradius}
\end{figure}

To our knowledge these results are unprecedented for the Einstein-Vlasov system. To close this section, we want to put our results in context with the fluid case, i.e., the Einstein-Euler system, since each isotropic steady state of the Einstein-Vlasov system corresponds to a steady state in the fluid case. It was shown in \cite{HaLiRe2020} that linear stability in the fluid setting implies linear stability in the Vlasov case. Furthermore, the linear stability  behavior  in the fluid case can be obtained by considering the mass turning points of the mass-radius curve \cite{HaLi}. It is therefore of interest whether the present equations of state provide multiple changes in stability in the fluid setting as well. However, from the results in \cite{HaLi} and Figure~\ref{img:massradius} it follows that this is not the case. Thus, the presence of multiple stability changes seems to be a pure feature of the Vlasov matter setting.  

%
%
\section{Observations and perspectives}\label{ssc:conclusion}
We conclude this work by mentioning certain observations, possible explanations, and further questions which we deem possibly connected to the stability  behavior obtained in Section~\ref{sc:results}. This section therefore involves a great deal of speculation and should be considered with caution.

Let us try to explain how one might make sense of the peculiar behavior observed in the previous section. We are certain that the different scales of values of the equation of state $\Phi_i$ are crucial for our results. Low energetic particles are thus assigned a much higher weight in comparison with higher energetic particles. These low energetic particles appear close to the origin and are the main driver of the dense core of the core-halo configurations. It seems as if the dense core has a stabilizing effect on the steady state as a whole.

How does this explain our observations? Firstly, the steady state gets stabilized for larger values of $z_c$ as the core gets more and more dense. For example, in Schwarzschild coordinates for  \mbox{$i=250$}, cf. Figure~\ref{img:aal}, the steady state is very close to collapsing for \mbox{$z_c = 0.517 $} as the ansatz function is equal to the case of \mbox{$k=1$} for \mbox{$z_c \in [0,0.437]$}, and since instability sets in at $z_c\approx 0.5$ for $k=1$. However, as the ansatz function increases rapidly, the dense core stabilizes the steady states, no collapse happens, and the family is stable up to values of \mbox{$z_c = 0.918 $}.

Secondly, the stabilizing effect can prevail even when instability has already set in. In maximal areal coordinates for \mbox{$i=270$} the steady states collapse for  \mbox{$ 0.509 \leq z_c \leq 0.698 $} after which the densification of the core has had such a strong stabilizing effect that the steady states are stable again for \mbox{$ 0.712 \leq z_c \leq 0.964 $}.

As we can apparently retrace the stability behavior to the low energetic particles at the core, it is natural to ask if a long spatial tail of the steady state can even affect the stability properties. For collapsing solutions only the very dense core collapses while the tail remains structurally unchanged. In the context of multi-shell solutions with multiple spatially separated shells, cf.\ \cite{AnRe2007}, it is quite obvious that the stability of the inner shell should be independent from the stability of the outer shells. 

If it is indeed the case that the stability behavior of the core is in some sense independent from that of the tail, we might have to find a local criterion for the stability properties. One canonical candidate is a local binding energy. We define it as 
\begin{equation*}
	E_b^{local}(r) = \frac{n(r)-m(r)}{n(r)},
\end{equation*}
where $n(r)$ and $m(r)$ are the number of particles and the mass contained inside the radius $r$. Outside of the radial support this obviously equals the binding energy $E_b$. In particular, for core-halo configurations the local binding energy appears appropriate, since the tail influences---and even dominates---the binding energy of the core for large values of $r$. We have studied the local binding energy for our equations of state, but we did not find an universal criterion for the onset of instability that can be constructed from such considerations.

We note that for the equations of state from Section~\ref{ssc:history}, e.g., the King model and polytropes, the local binding energy $E_b^{local}$ is monotonically increasing in $r$ for fixed values of $z_c$. 

Besides the local binding energy, it is possible to consider a modified binding energy that takes into account the associated energy-Casimir functional that is extremized by the steady state, cf. \cite{HaRe2013}. One could for example replace $N$ with the Casimir functional. This and further attempts were to no avail.

We have also tried to find other curves or special values that might be connected to the stability of isotropic steady states. Turning points of the mass-radius spiral, which determine stability behavior for the Einstein-Euler system \cite{HaLi}, do not determine the stability in the Einstein-Vlasov case as earlier observed in \cite{Praktikum20} and as mentioned above. Furthermore, there does not seem to exist a universal threshold value of $z_c$ or  $\frac{2M}{R}$ or  \mbox{$\sup_{r>0} \frac{2m}{r}$} such that isotropic steady states are always unstable for larger values of the respective quantities. 

\section{Conclusion}\label{sc:realconclusion}

In this work, we have numerically investigated stability issues for collisionless equilibria in general relativity, more precisely, stability of stationary solutions of the spherically symmetric, asymptotically flat Einstein-Vlasov system in Schwarzschild and in maximal areal coordinates. The dynamic stability of isotropic steady states of said system has attracted a lot of interest in both physics and mathematics communities as reviewed in Section~\ref{ssc:history}. 

In Section~\ref{ssc:evidencebinding}, we present convincing numerical evidence that the strong binding energy hypothesis does not hold true in general, i.e., the conjecture that along a one-parameter family of stationary solutions instability sets in precisely at the first local maximum of the binding energy. To do so we consider a family of isotropic equilibria induced by a fixed piecewise linear microscopic equation of state parametrized by their central-redshift value, see Section~\ref{sc:ansatzfunction}. The dynamic stability properties for several such equations of state are summarized in Figure~\ref{img:aal}. Notably, steady states seem to always be stable up to the first maximum of the binding energy, which has been conjectured and partially been proven to be the case in the past.
Furthermore, in Section~\ref{ssc:stabilitychanges}, we provide numerical examples showing that there may exist multiple changes from stability to instability along a one-parameter family of stationary solutions corresponding to the same microscopic equation of state. This constitutes an observation which is unprecedented in the Vlasov matter case.

All our numerical findings demonstrate that stability issues for the Einstein-Vlasov system may be more delicate than previously thought. In particular, we could not find any global quantity---such as the binding energy---from which it can be generally predicted whether or not an equilibrium is stable. In addition, it is unclear what is the mechanism behind the fact that the strong binding energy hypothesis holds for the various models mentioned in Section~\ref{ssc:history}. This work thus opens up exciting new questions, and the authors will be pleased if their numerical simulations prove to be of aid in order to gain more understanding of stability issues in general relativity.


\end{document}